# Correlation between the spin Hall angle and the structural phases of early 5*d* transition metals

Jun Liu, Tadakatsu Ohkubo, Seiji Mitani, Kazuhiro Hono and Masamitsu Hayashi[*]

*National Institute for Materials Science, Tsukuba 305-0047, Japan*

We have studied the relationship between the structure and the spin Hall angle of the early 5*d* transition metals in X/CoFeB/MgO (X=Hf, Ta, W, Re) heterostructures. Spin Hall magnetoresistance (SMR) is used to characterize the spin Hall angle of the heavy metals. Transmission electron microscopy images show that all underlayers are amorphous-like when their thicknesses are small, however, crystalline phases emerge as the thickness is increased for certain elements. We find that the heavy metal layer thickness dependence of the SMR reflects these changes in structure. The spin Hall angle largest $|\theta_{SH}|$ of Hf, Ta, W and Re (~0.11, 0.10, 0.23 and 0.07, respectively) is found when the dominant phase is amorphous-like. We find that the amorphous-like phase not only possesses large resistivity but also exhibits sizeable spin Hall conductivity, which both contribute to the emergence of the large spin Hall angle.

*Email: hayashi.masamitsu@nims.go.jp

Heavy metals with strong spin orbit coupling are attracting great interest recently[1-3] as such elements can generate significant amount of spin current via the spin Hall effect[4]. In heterostructures that contain a heavy metal layer and a magnetic layer, the spin current generated within the heavy metal layer can diffuse into the magnetic layer and trigger current induced magnetization switching[2, 5] and domain wall motion[6-9]. Key to the development of technologies that make use of such current induced effects is the large spin Hall angle of the heavy metal layer[10, 11].

The spin Hall angle, which defines the amount of spin current generated within the layer, is known to be element specific. First principle calculations[12] show that the spin Hall angle of the transition metals depends on the electron filling of the $d$-orbitals if the mechanism behind it is due to an intrinsic origin (i.e., band structure related)[13]. In addition, the spin Hall effect can originate from spin dependent scattering of electrons, commonly referred to as an extrinsic effect[14, 15]. Recently large spin Hall angle has been reported for the heavy transition metals[2, 16-19] and/or its alloy[20]. The large spin Hall angle for the early $5d$ transition metals (e.g. Ta and W) has been associated[2, 16, 21] with the appearance of the distorted tetragonal phase, often known as the A-15 β-phase[22-25]. In order to develop viable devices utilizing the spin Hall effect, it is thus essential to understand how the structure of the heavy metal influences, if any, the size of the spin Hall angle. Identifying the relationship between the structural phase of the heavy metal layer and its spin Hall angle, however, requires thorough and systematic studies.

Here we show studies on the spin Hall angle of the early $5d$ transition metals (Hf, Ta, W, Re) in magnetic heterostructures. Spin Hall magnetoresistance (SMR)[18, 26-28] and transmission electron microscopy (TEM) are used to evaluate, respectively, the spin transport properties and the structural phase of the films. We find that the heavy metal layer thickness dependence of the SMR reflects changes in the structure of the heavy metal layer. The spin Hall angle is estimated for various structural phases emerging in each element and we find that phases with larger disorder give rise to larger spin Hall angle.

Films are deposited on Si substrates with 100 nm thick thermal oxide. The film structure is Sub.|$d$ X|1 CoFeB|2 MgO|1 Ta (units in nm) and the composition of the CoFeB sputtering target is Co:Fe:B=20:60:20 at%. The underlayer X is one of the early 5$d$ transition metals (X=Hf, Ta, W, Re). All films are annealed in vacuum at 300 °C for 1 hour. Cross sectional high resolution transmission electron microscopy (HRTEM) images are obtained using Taitan G2 80-200. For the TEM studies, films are prepared using focused ion (Ga$^+$) beam. Resistivity and the spin Hall magnetoresistance are studied using patterned structures that enable four point probe resistance measurements. The thickness $d$ of the underlayer X is linearly varied across the substrate to allow resistance measurements of samples with different $d$ with an otherwise same film structure.

The underlayer thickness dependence of the spin Hall magnetoresistance is shown in Figs. 1(b-e) for all structures studied[29]. The spin Hall magnetoresistance (SMR) is estimated using the difference in resistance when the magnetization lies along the $y$ ($R_{XX}^Y$) and the $z$ ($R_{XX}^Z$) axes (see Ref. 28 for the details of the experimental method). The SMR is defined as $\left(R_{XX}^Y - R_{XX}^Z\right)/R_{XX}^Z$. Definition of the coordinate axes is shown in Fig. 1(a). For all structures, SMR takes a maximum at a given heavy metal layer thickness. According to a model that describes SMR using a spin-diffusion approximation in the heavy metal layer[28, 30], the SMR peak value is proportional to the square of the spin Hall angle whereas the heavy metal layer thickness at which the peak takes place gives information on the spin diffusion length. The heavy metal layer thickness ($d$) dependence of the SMR is fitted with an analytical formula[28, 30, 31]:

$$\frac{\Delta R_{XX}}{R_{XX}^{Z}} \sim \theta_{SH}^{2} \frac{\lambda_{N}}{d} \frac{\tanh(d/2\lambda_{N})}{1+\xi} \left[1 - \frac{1}{\cosh(d/\lambda_{N})}\right] \quad (1)$$

where $\theta_{SH}$ and $\lambda_{N}$ are the spin Hall angle and spin diffusion length, respectively, of the heavy metal layer. $\xi \equiv (\rho_{N} t_{F}/\rho_{F} d)$ describes the current shunting effect into the magnetic layer, where $\rho_{F}$ and $t_{F}$ represent the resistivity and the thickness of the magnetic layer and $\rho_{N}$ is the resistivity of the heavy metal layer. $\theta_{SH}$ and $\lambda_{N}$ are used as the fitting parameters and we use $\rho_{F} \sim 160$ μΩ·cm for CoFeB and experimentally determined $\rho_{N}$ (Figs. 1(f)-1(k)) for each heavy metal layer.[9] The fitted curves are shown by the solid lines in Figs. 1(b-e).

As will be evident from the TEM analysis presented below, the heavy metal layer thickness dependence of the SMR reflects changes in the structures of the heavy metal layer. For the Hf underlayer films (Fig. 1(b)), we find two peaks in SMR vs. $d$, which indicate the presence of two different (structural) phases that contribute to the SMR. We thus fit the results with two sets of parameters: one that fits the thin Hf underlayer films ($d<\sim 2$ nm) and the other that fits the thicker films ($d>\sim 2$ nm). Fit to the experimental data for the Ta underlayer films (Fig. 1(c)) shows good agreement for the entire thickness range studied. In contrast, the SMR for the W underlayer films (Fig. 1(d)) deviates from the fitting when $d$ is larger than ~5 nm (the reason for this is discussed below). As for the Re underlayer films (Fig. 1(e)), we find that the roughness of the Re underlayer films are considerably larger than that of the other films. This is likely to do with the poor wetting of Re on $SiO_2$. To overcome this issue, a thin (0.5 nm thick) Ta seed layer is formed before depositing the Re underlayer. By adding the thin Ta seed layer, we find that the surface roughness significantly improves. The Re layer thickness dependences of SMR for both heterostructures, i.e. with and without the Ta seed layer, are shown in Fig. 1(e). The fitting results are summarized in Table 1.

The inverse of the film sheet resistance is plotted against the heavy metal layer thickness $d$ in Figures 1(f-i). The sheet resistance is obtained by multiplying the film resistance ($R_{XX}^Z$) with the width ($w$~10 μm) and the inverse of length ($L$~25 μm) of the device. If the inverse of the sheet resistance scales linearly with $d$, the slope gives the resistivity ($\rho_N$) of the heavy metal layer. The solid lines in Figs. 1(f-i) show linear fits to the data in appropriate ranges. For W (Fig. 1(j)) and Re (Fig. 1(k)) underlayer films, the sheet resistance inverse shows a jump at $d$~5 and ~6 nm, respectively. The jump occurs for both Re underlayer films with and without the Ta seed layer. The thickness at which the sheet resistance inverse shows a discontinuity is defined as $d_C$, hereafter. For all structures, the slope of the sheet resistance inverse vs. $d$ changes when $d$ approaches zero. The degree of deviation is larger for Hf (Fig. 1(f)) and Re (no Ta seed, Fig. 1(i)) underlayer films. For these heterostructures, we estimate the resistivity of the thinner regime too. The resistivity estimated from the linear fitting is summarized in Table 1.

Figure 2 shows cross-sectional TEM images of one of the Hf underlayer films ($d$~6 nm). The dark layer in the bright field TEM (BFTEM) image (Fig. 2(a)) corresponds to the Hf underlayer and shows that it grows uniformly across the substrate with small roughness. The MgO layer is (001) textured, as observed in the HRTEM image (Fig. 2(b)). Lattice fringes of the (001) textured MgO layer have been observed for all the heterostructures studied. From the HRTEM image, we find that the Hf layer seems to be partially crystallized: the bottom ~2 nm thick Hf shows amorphous feature whereas the top ~4 nm thick Hf is crystalline. To identify its structure, nano-beam diffraction patterns corresponding to the lower and upper parts of the Hf layer, indicated by regions C and B in Fig. 2(b), are shown in Figs. 2(d) and 2(e), respectively. The ring like diffraction pattern for the lower part of the Hf layer (Fig. 2(e)) verifies that its structure is amorphous whereas the upper part of the pattern indicates that the Hf has an hcp structure. The diffraction pattern of the CoFeB layer (Fig. 2(c)) shows a halo ring like structure, suggesting that the CoFeB layer is amorphous. In contrast to Hf, the Ta underlayer shows no sign of crystallization even when its thickness is ~7 nm (Figs. 2(f) and 2(g)).

HRTEM images of the W underlayer films (Figs. 3(a) and 3(c)) show that the structure of the W underlayer is different depending on its thickness. The thinner W underlayer ($d$~3 nm) possesses amorphous-like structure whereas the thicker W ($d$~6 nm) clearly shows crystalline lattices. Nano-beam diffraction pattern of the thinner W underlayer (Fig. 3(b)) shows halo rings with a few weak diffraction spots corresponding a bcc lattice. This suggests that the thin W underlayer (~3 nm thick) is predominantly amorphous but is mixed with a bcc phase. In contrast, the nano-beam diffraction pattern taken from the thicker W underlayer (Fig. 3(d)) indicates that the W forms a fully crystallized bcc structure.

Figures 4(a,b,d,e) show cross-sectional TEM images of representative Re underlayer ($d$~4) films with and without the Ta seed layer. The BFTEM images show that the roughness of the Re layer is significantly improved by insertion of the Ta seed layer (Figs. 4(a) and 4(d)). Interestingly, the structure of the Re layer is also different for the two films. The HRTEM (Fig. 4(b)) and the corresponding nano-beam diffraction pattern (Fig. 4(c)) indicate that the Re underlayer without the Ta seed layer is a mixture of amorphous and hcp phases. In contrast, the Re underlayer with the Ta seed layer is predominantly amorphous (Fig. 4(e) and 4(f)) with little sign of the crystalline phase. When the Re underlayer thickness is increased beyond $d_C$, it displays a fully crystallized structure. The BFTEM image, shown in Fig. 4(g), again illustrates the large roughness of the Re layer without the Ta seed layer. However, the HRTEM (Fig. 4(h)) and the nanobeam diffraction pattern (Fig. 4(i)) display the crystalline hcp structure of the Re layer.

We next discuss the relationship between the spin Hall angle estimated from the SMR measurements and the structural phases of the heavy metal layer found from the TEM studies. Table 1 summarizes the results from both measurements. The two peaks found in the SMR vs. $d$ for the Hf underlayer films (Fig. 1(b)) correspond to the SMR signals from the thin amorphous and the thicker hcp phases of Hf. Note that it is the difference in the spin diffusion length of the two phases that causes the appearance of the two SMR peaks. The estimated spin Hall angle for the amorphous phase is larger than that of the hcp phase. For the Ta underlayer films, the structure remains amorphous throughout the thickness studied. For these films, the experimentally obtained SMR values can be fitted well with a single curve (Fig. 1(c)) and the inverse sheet resistance varies linearly with the Ta layer thickness (Fig. 1(g)).

For the W underlayer films, the jump in the inverse sheet resistance at $d=d_C$ (~5 nm) (Fig. 1(j)) corresponds to the structural transition[32, 33] of W between the amorphous phase and the crystalline bcc phase[9, 16]. When the underlayer thickness is larger than $d_C$, the SMR (Fig. 1(c)) drops considerably and the experimental data can no longer be fitted with Eq. (1), suggesting that the spin Hall angle is significantly reduced for the bcc W underlayer[16]. This also applies to the Re underlayer films: the TEM images of the Re underlayer films show that Re undergoes a structural phase transition at $d=d_C$ (~6 nm) from a predominantly amorphous phase to a highly crystalline hcp phase. For the Re underlayer films with the Ta seed layer (Fig. 1(d)), the SMR drops beyond $d=d_C$ (it is difficult to discuss the changes in SMR for films without the Ta seed layer), again indicating that the crystalline hcp phase has a smaller spin Hall angle than the amorphous phase. Note that the spin Hall angle of the amorphous W is smaller than that reported previously by Pai *et al.*[16] which may partly be because the thin W here is mixed with the bcc phase, as evident from Fig. 3(b). Similarly, the spin Hall angle for the Re underlayer films with the Ta seed layer is larger than that without the Ta seed layer. This may reflect the degree of disorder: the former exhibits a stronger halo like diffraction pattern indicating that its structure contains lesser hcp phase than the latter.

The spin Hall conductivity ($\sigma_{SH}$) can be estimated using a phenomenological relationship: $\theta_{SH} \sim \rho_N \cdot \sigma_{SH}$. The value of each element is summarized in Table 1. These results show that the relatively large spin Hall angle found in the early transition metals is due to two factors. First, the large resistivity (>~100 µΩ·cm) of the elements, which mostly originates from the amorphous like structure, defines the frequency of the scattering events and is essential in obtaining large $\theta_{SH}$. The second factor is the relatively large spin Hall conductivity, which may be related to the filling of the $5d$ orbitals[12, 17, 34]. However, the reason behind the large spin Hall conductivity for metals with amorphous like disordered structure is not clear and will require theoretical investigations. Note that the estimated $\sigma_{SH}$ for Ta (~0.7×10$^3$ Ω$^{-1}$cm$^{-1}$) and W (~1.8×10$^3$ Ω$^{-1}$cm$^{-1}$) are much larger than those found in an earlier report[10].

In summary, we have studied the relationship between the structure and the spin Hall angle of the early $5d$ transition metals (X=Hf, Ta, W, Re) in X/CoFeB/MgO heterostructures. We find that the heavy metal layer thickness dependence of the spin Hall magnetoresistance (SMR) reflects changes, if any, in the structure of the heavy metal. The spin Hall angle ($\theta_{SH}$) is estimated from the SMR for various phases of each element and we find the degree of structural disorder influences $\theta_{SH}$. All elements studied show the largest |$\theta_{SH}$| when the dominant phase is amorphous-like. Using simple estimates, we find that not only the large resistivity but also the sizeable spin Hall conductivity contribute to the large spin Hall angle found for the amorphous phases.. These results provide insight into the origin of the spin Hall effect in transition metals that are technologically important, in particular, for devices that utilize spin orbit effects.

**Acknowledgements**

We thank S. Takahashi for valuable discussion on the theory of spin Hall magnetoresistance. This work was partly supported by MEXT R & D Next-Generation Information Technology, the JSPS KAKENHI Grant Numbers 23506017, 15H05702.

**Figure captions:**

**Figure 1.** (a) Schematic illustration of the device measured and the definition of the coordinate axes. (b-e) The heavy metal layer thickness dependence of the spin Hall magnetoresistance ($\Delta R_{XX}/R_{XX}^Z$). Solid and dashed lines show the fitting results using Eq. (1) for appropriate thickness ranges. (f-i) The inverse of the sheet resistance [$1/R_{XX}^Z/(w/L)$] versus the heavy metal layer thickness. (j, k) Expanded view of the plots shown in (h) and (i). Solid and dashed lines in (f-k) represent linear fitting to the data for appropriate thickness ranges. The heavy metal layer is Hf (b, f), Ta (c, g), W (d, h, j) and Re (e, i, k). The open symbols in (e, i, k) show results from Re underlayer films with a 0.5 nm thick Ta seed layer.

**Figure 2.** (a,b) Cross sectional bright field TEM (a) and high resolution TEM (b) images of Sub.|6 Hf|1 CoFeB|2 MgO|1 Ta. (c-e) Nano-beam diffraction patterns of regions A (c), B (d) and C (e) marked in (b). (f) Cross sectional high resolution TEM image of Sub.|7 Ta|1 CoFeB|2 MgO|1 Ta. (g) Nano-beam diffraction pattern of region D indicated in (f).

**Figure 3.** (a) Cross sectional high resolution TEM image of Sub.|3 W|1 CoFeB|2 MgO|1 Ta. (b) Nano-beam diffraction pattern of region A indicated in (a). (c) Cross sectional high resolution TEM (e) image of Sub.|6 W|1 CoFeB|2 MgO|1 Ta. (d) Nano-beam diffraction pattern of region B indicated in (c).

**Figure 4.** (a,b) Cross sectional bright field TEM (a) and high resolution TEM (b) images of Sub.|4 Re|1 CoFeB|2 MgO|1 Ta. (c) Nano-beam diffraction pattern of region A indicated in (b). (d,e) Cross sectional bright field TEM (d) and high resolution TEM (e) images of Sub.|0.5 Ta|4 Re|1 CoFeB|2 MgO|1 Ta. (f) Nano-beam diffraction pattern of region B indicated in (e). (g,h) Cross sectional bright field TEM (g) and high resolution TEM (h) images of Sub.|10 Re|1 CoFeB|2 MgO|1 Ta. (i) Nano-beam diffraction pattern of region C indicated in (h).

**Table 1.** The resistivity ($\rho_N$), the spin diffusion length ($\lambda_N$), the absolute values of the spin Hall angle ($|\theta_{SH}|$) and the spin Hall conductivity ($|\sigma_{SH}|$) of the early 5$d$ transition metals formed in magnetic heterostructures.

| Heavy metal [a] | $\rho_N$ μΩcm | $\lambda_N$ nm | $|\theta_{SH}|$ | $\sigma_{SH} \times 10^3$ $\Omega^{-1}$cm$^{-1}$ |
|---|---|---|---|---|
| Hf (a) | 406 | 0.3 | 0.11 | 0.28 |
| Hf (hcp) | 155 | 1.3 | 0.07 | 0.47 |
| Ta (a) | 159 | 0.4 | 0.10 | 0.65 |
| W (a+bcc) | 125 | 1.3 | 0.23 | 1.82 |
| W (bcc) | 12 | N/A | N/A | N/A |
| 0.5 Ta|Re (a) | 172 | 1.0 | 0.07 | 0.42 |
| Re (a+hcp) | 102 (~331) | 1.5 | 0.04 | 0.37 |
| Re (hcp) | 28 | N/A | N/A | N/A |

[a] a: amorphous, bcc: body centered cubic, hcp: hexagonal close packed

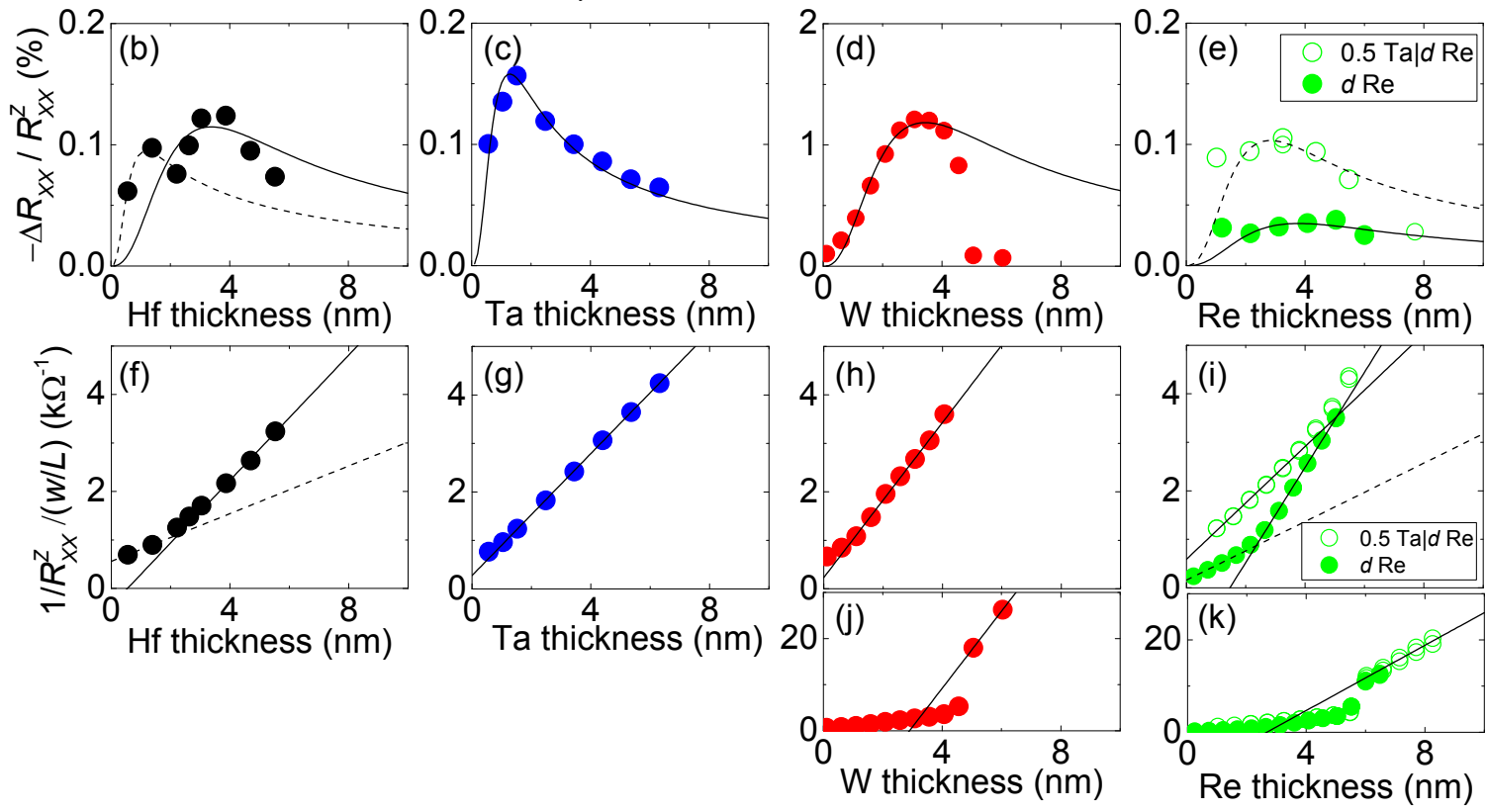

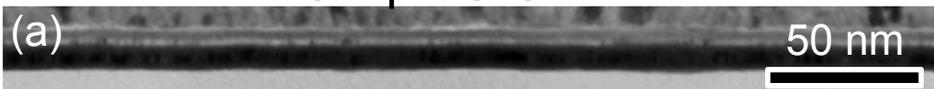

**6 Hf|1 CoFeB**

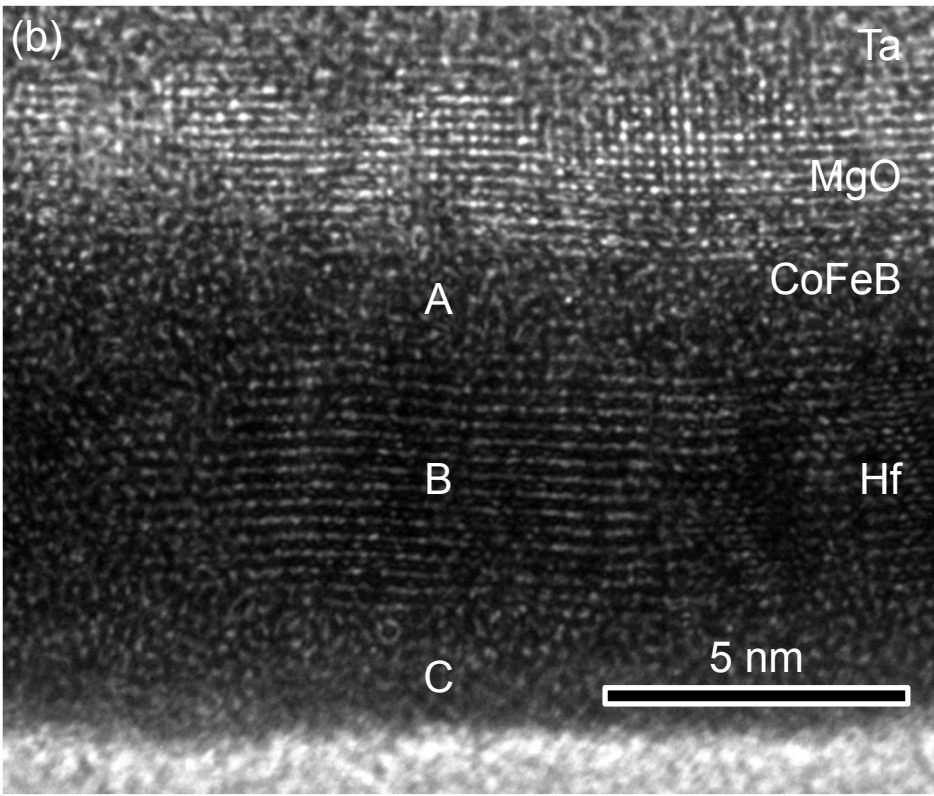

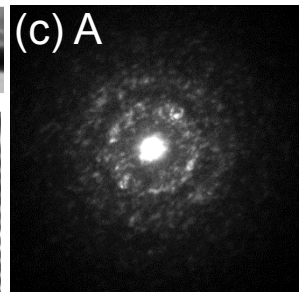

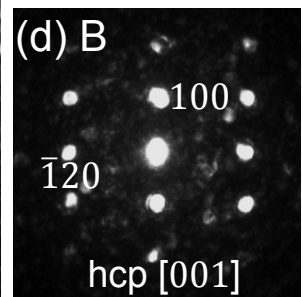

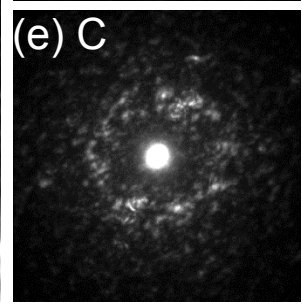

**7 Ta|1 CoFeB**

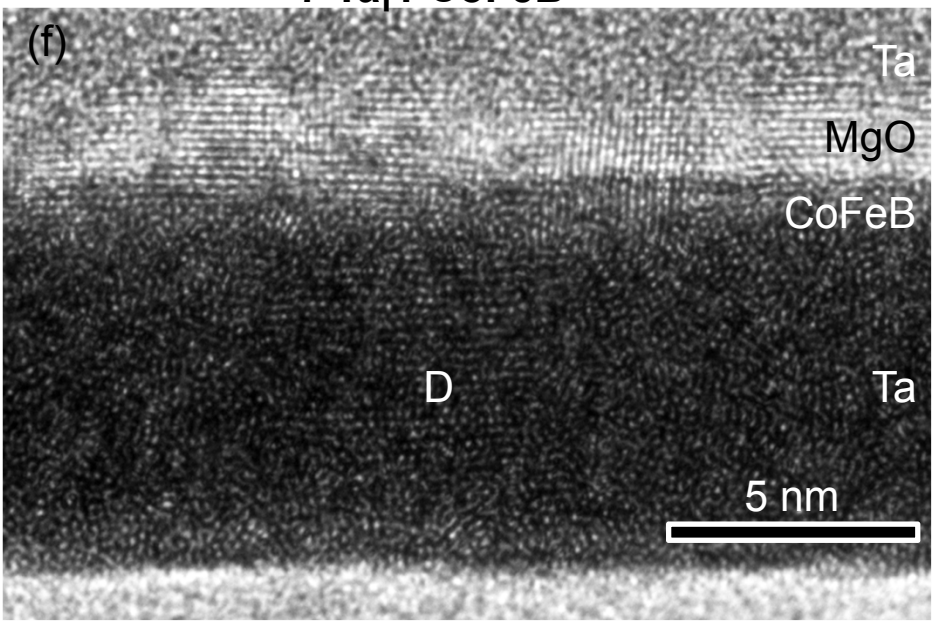

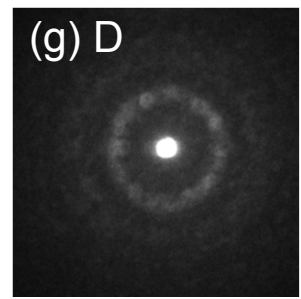

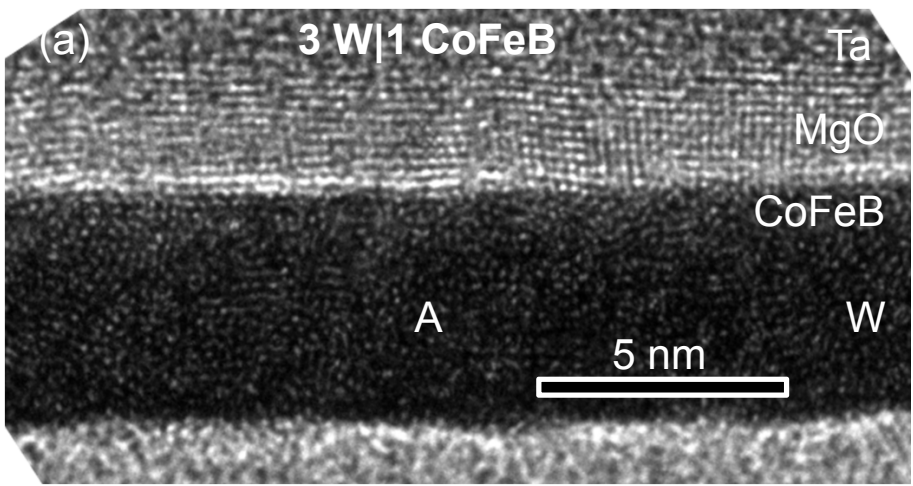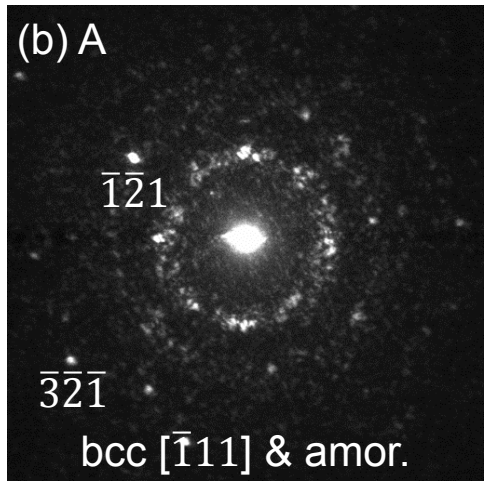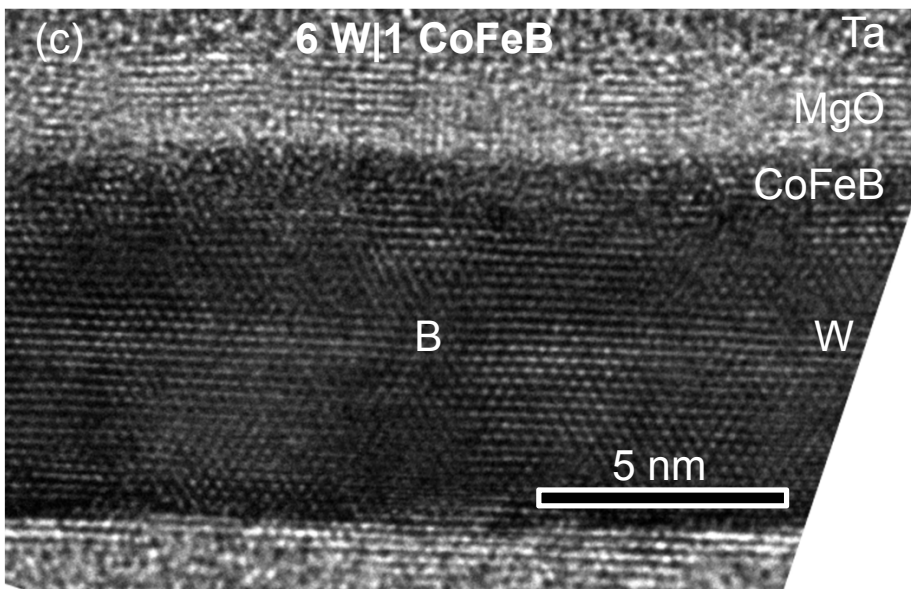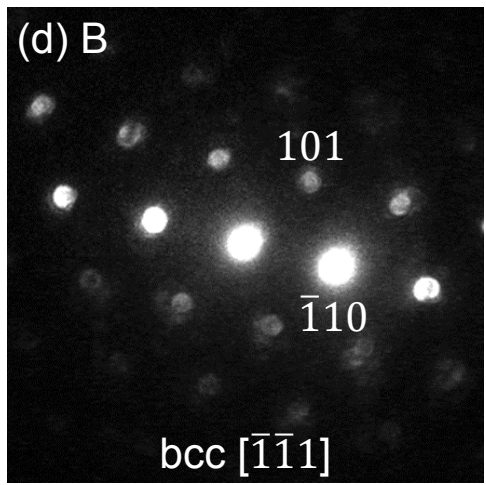

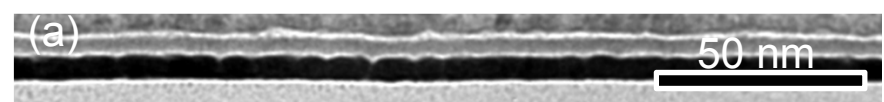
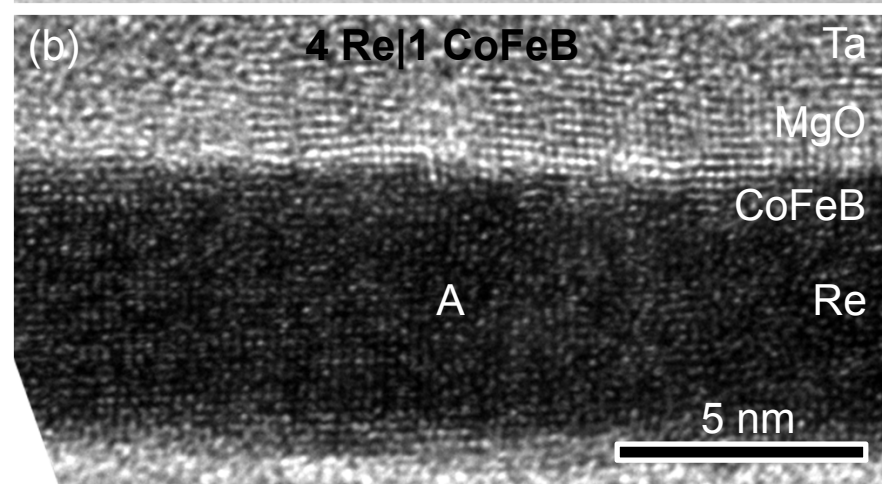
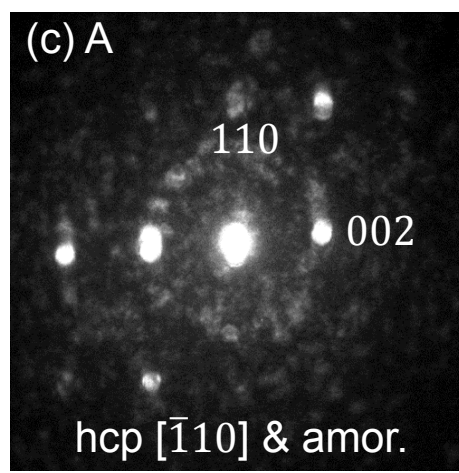
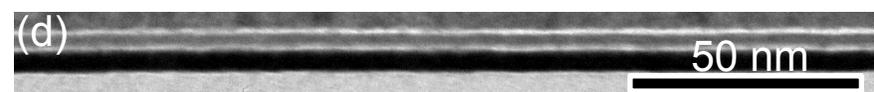
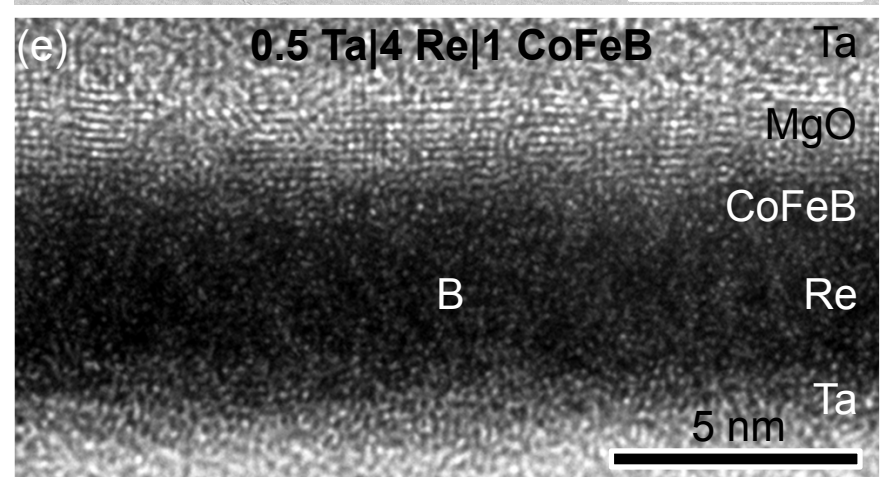
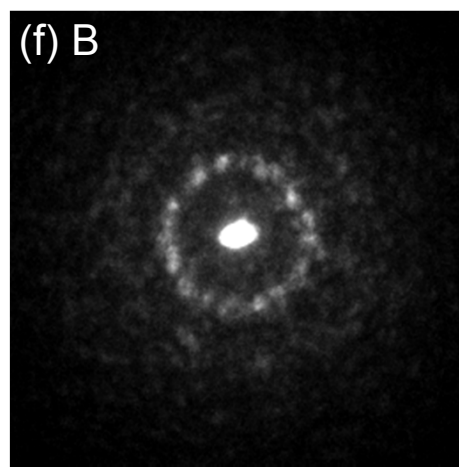
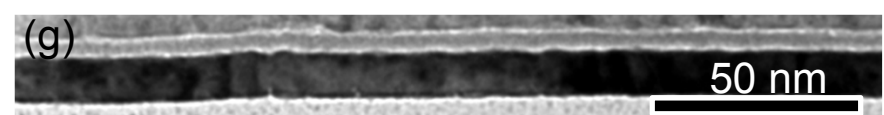
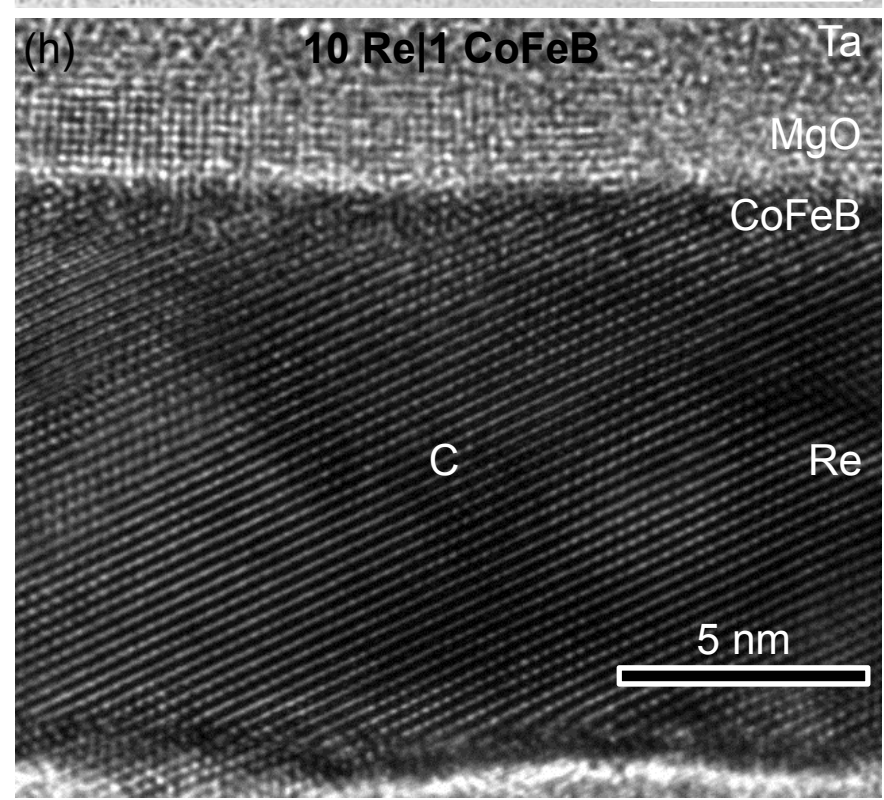
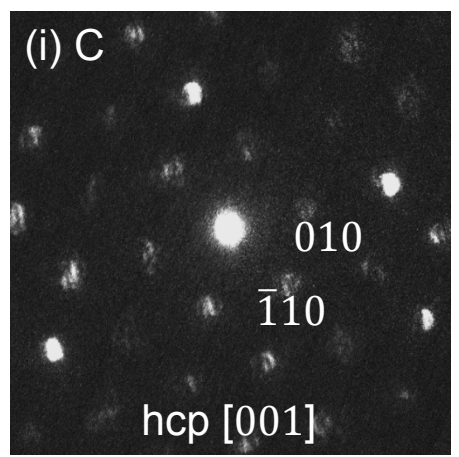

# Supplementary material for

# Correlation between the spin Hall angle and the structural phases of early 5d transition metals

Jun Liu, Tadakatsu Ohkubo, Seiji Mitani, Kazuhiro Hono and Masamitsu Hayashi*

*National Institute for Materials Science, Tsukuba 305-0047, Japan*

## I. Derivation of Eq. (1)

Equation (1) in the main text is derived from Eq. (22) of Chen et al.,[1] which reads:

$$\frac{\Delta R_{XX}}{R_{XX}^Z} = \theta_{SH}^2 \frac{\lambda_N}{d} \text{Re}\left[\frac{2\rho_N \lambda_N G \tanh^2(d/2\lambda_N)}{1 + 2\rho_N \lambda_N G \coth(d/\lambda_N)}\right] \tag{S1}$$

where $\theta_{SH}$, $\lambda_N$, $\rho_N$ and $d$, are the spin Hall angle, the spin diffusion length, the resistivity and the thickness of the heavy metal layer, respectively. $G$ is the spin mixing conductance defined at the interface of the heavy metal (HM) and the ferromagnetic insulator (FI). Rearranging Eq. (S1) results in

$$\frac{\Delta R_{XX}}{R_{XX}^Z} = \theta_{SH}^2 \frac{\lambda_N}{d} \tanh(d/2\lambda_N)\left[1 - \frac{1}{\cosh(d/\lambda_N)}\right]\text{Re}\left[\frac{g}{1+g}\right] \tag{S2}$$

$$g \equiv 2\rho_N \lambda_N G \coth(d/\lambda_N)$$

Here we have used the following relation: $\tanh(x)\tanh(x/2) = 1 - 1/\cosh(x)$. In the limit of $|\text{Im}[G]| \ll |\text{Re}[G]|$, Eq. (S2) is reduced to

$$\frac{\Delta R_{XX}}{R_{XX}^Z} = \theta_{SH}^2 \frac{\lambda_N}{d} \tanh(d/2\lambda_N)\left[1 - \frac{1}{\cosh(d/\lambda_N)}\right]\frac{\text{Re}[g]}{1+\text{Re}[g]} \tag{S3}$$

Equation (S3), which is the same with Eq. (27) of Chen et al., further reduces when we assume $|\text{Re}[G]| \gg 1$, that is

$$\frac{\Delta R_{XX}}{R_{XX}^Z} = \theta_{SH}^2 \frac{\lambda_N}{d} \tanh(d/2\lambda_N)\left[1 - \frac{1}{\cosh(d/\lambda_N)}\right] \tag{S4}$$

Equation (S4), identical to Eq. (29) of Chen *et al.*, can be applied to a HM/FI bilayer system, in which the current flows only through the heavy metal layer. In metallic bilayers where the FI is replaced with a ferromagnetic metal (FM), current flows not only through the HM layer but also through the FM layer. In order to take into account such current shunting effect, we consider a circuit shown in Fig. S1. The resistances of the HM and FM layers are defined as $R_{HM}$ and $R_{FM}$, respectively, and the resistance change caused by the spin Hall magnetoresistance is represented by $\Delta R$.

In the absence and presence of SMR, the total resistance of the circuit reads, respectively:

$$R_0 = \frac{R_{HM} R_{FM}}{R_{HM} + R_{FM}} \tag{S5a}$$

$$R_{SMR} = \frac{(R_{HM} + \Delta R_{SMR}) R_{FM}}{(R_{HM} + \Delta R_{SMR}) + R_{FM}} \tag{S5a}$$

The magnetoresistance (SMR) for the HM/FM system can be calculated as the following:

$$\left.\frac{R_0 - R_{SMR}}{R_0}\right|_{HM/FM} = \frac{\Delta R_{SMR}}{R_{HM}\left[1 + \frac{R_{HM} + \Delta R_{SMR}}{R_{FM}}\right]} \approx \frac{\Delta R_{SMR}}{R_{HM}\left[1 + \frac{R_{HM}}{R_{FM}}\right]} \tag{S6}$$

The second nearly-equal sign is obtained by assuming $R_{HM} \gg \Delta R_{SMR}$. In the HM/FI system, one can substitute $R_{FM} \to \infty$ into Eq. (S6) and obtain:

$$\left.\frac{R_0 - R_{SMR}}{R_0}\right|_{HM/FI} \approx \frac{\Delta R_{SMR}}{R_{HM}} \tag{S7}$$

Equation (S6) thus can be rewritten as:

$$\left.\frac{R_0 - R_{SMR}}{R_0}\right|_{HM/FM} \approx \left.\frac{R_0 - R_{SMR}}{R_0}\right|_{HM/FI} \times \frac{1}{1 + R_{HM}/R_{FM}} \tag{S8}$$

Equation (S8) shows that in order to apply Eq. (S4) for the HM/FM system, one needs to multiply Eq. (S4) with $\frac{1}{1 + R_{HM}/R_{FM}}$, i.e.

$$\left.\frac{\Delta R_{XX}}{R_{XX}^Z}\right|_{HM/FM} = \theta_{SH}^2 \frac{\lambda_N}{d} \frac{\tanh(d/2\lambda_N)}{1 + \xi}\left[1 - \frac{1}{\cosh(d/\lambda_N)}\right] \tag{S9}$$

where $\xi \equiv (\rho_N t_F / \rho_F d)$. $\rho_F$ and $t_F$ represent the resistivity and the thickness of the magnetic layer, respectively. Equation (S9) is identical to Eq. (1) in the main text.

**Figure captions**

**Figure S1.** Equivalent circuit to describe the spin Hall magnetoresistance in a heavy metal (HM)/ferromagnetic metal (FM) bilayer. The resistances of the HM and FM layers are $R_{HM}$ and $R_{FM}$, respectively, and the resistance change caused by the spin Hall magnetoresistance is defined as $\Delta R$.

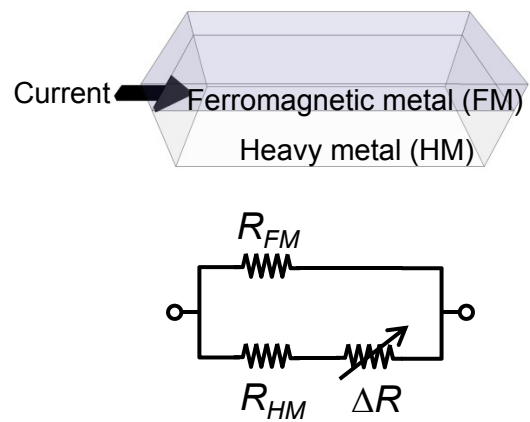

Fig. S1